\documentclass[pra,twocolumn,superscriptaddress,nobibnotes,letterpaper]{revtex4-2}

\usepackage[pdftex]{graphicx}
\usepackage[pdftex]{epsfig}
\usepackage{amsmath}
\usepackage{amssymb}
\usepackage{amsfonts}
\usepackage{color}
\usepackage{wrapfig}
\usepackage{eucal}
\usepackage{hhline}
\usepackage{threeparttable}
\usepackage{supertabular}
\usepackage{multirow}
\usepackage{tabularx}
\usepackage[separate-uncertainty=true]{siunitx}
\usepackage{nicefrac}

\usepackage{soul}
\usepackage{comment}

\usepackage{xr}
\usepackage{array} % make column bigger (vertical)
\usepackage[colorlinks=true, allcolors=blue]{hyperref} %hyperrefs

\newcommand{\ket}[1]{\left|#1\right\rangle}
\newcommand{\expec}[1]{\left\langle #1\right\rangle}

\newcommand{\opt}{\mathrm{o}}
\newcommand{\AB}{\mathrm{AB}}
\newcommand{\inp}{\mathrm{in}}
\newcommand{\outp}{\mathrm{out}}
\newcommand{\ro}{\mathrm{out}}
\newcommand{\order}[1]{\mathcal{O}(#1)}
\newcommand{\pb}{p_{\mathrm{b}}}
\newcommand{\pr}{p_{\mathrm{r}}}

\newcommand{\nth}{n_\mathrm{th}}

\newcommand{\und}[1]{_\textrm{#1}}

\begin{document}

\title{Optomechanical quantum teleportation}\thanks{This work was published in \href{https://doi.org/10.1038/s41566-021-00866-z}{Nature Photon.\ \textbf{15}, 817--821 (2021).}}

\author{Niccol\`{o} Fiaschi}\thanks{These authors contributed equally to this work.}
\affiliation{Kavli Institute of Nanoscience, Department of Quantum Nanoscience, Delft University of Technology, 2628CJ Delft, The Netherlands}

\author{Bas Hensen}\thanks{These authors contributed equally to this work.}
\affiliation{Kavli Institute of Nanoscience, Department of Quantum Nanoscience, Delft University of Technology, 2628CJ Delft, The Netherlands}

\author{Andreas Wallucks}
\affiliation{Kavli Institute of Nanoscience, Department of Quantum Nanoscience, Delft University of Technology, 2628CJ Delft, The Netherlands}

\author{Rodrigo Benevides}
\affiliation{Kavli Institute of Nanoscience, Department of Quantum Nanoscience, Delft University of Technology, 2628CJ Delft, The Netherlands}
\affiliation{Photonics Research Center, Applied Physics Department, Gleb Wataghin Physics Institute, P.O.\ Box 6165, University of Campinas -- UNICAMP, 13083-970 Campinas, SP, Brazil}

\author{Jie Li}
\affiliation{Kavli Institute of Nanoscience, Department of Quantum Nanoscience, Delft University of Technology, 2628CJ Delft, The Netherlands}
\affiliation{Zhejiang Province Key Laboratory of Quantum Technology and Device, Department of Physics, Zhejiang University, Hangzhou 310027, China}

\author{Thiago P.\ Mayer Alegre}
\affiliation{Photonics Research Center, Applied Physics Department, Gleb Wataghin Physics Institute, P.O.\ Box 6165, University of Campinas -- UNICAMP, 13083-970 Campinas, SP, Brazil}

\author{Simon Gr\"oblacher}
\email{s.groeblacher@tudelft.nl}
\affiliation{Kavli Institute of Nanoscience, Department of Quantum Nanoscience, Delft University of Technology, 2628CJ Delft, The Netherlands}

\begin{abstract}
Quantum teleportation, the faithful transfer of an unknown input state onto a remote quantum system~\cite{Bennett1993}, is a key component in long distance quantum communication protocols~\cite{Sangouard2011} and distributed quantum computing~\cite{Raussendorf2001,Barz2012}. At the same time, high frequency nano-optomechanical systems~\cite{Chan2011} hold great promise as nodes in a future quantum network~\cite{Kimble2008}, operating on-chip at low-loss optical telecom wavelengths with long mechanical lifetimes. Recent demonstrations include entanglement between two resonators~\cite{Riedinger2018}, a quantum memory~\cite{Wallucks2020} and microwave to optics transduction~\cite{Forsch2020,Jiang2020,Mirhosseini2020}. Despite these successes, quantum teleportation of an optical input state onto a long-lived optomechanical memory is an outstanding challenge. Here we demonstrate quantum teleportation of a polarization-encoded optical input state onto the joint state of a pair of nanomechanical resonators. Our protocol also allows for the first time to store and retrieve an arbitrary qubit state onto a dual-rail encoded optomechanical quantum memory. This work demonstrates the full functionality of a single quantum repeater node, and presents a key milestone towards applications of optomechanical systems as quantum network nodes.
\end{abstract}

\maketitle

 High frequency nano-optomechanical systems~\cite{Chan2011}, besides their appeal for probing fundamental quantum physics~\cite{Aspelmeyer2014}, also hold great promise as nodes in a future quantum network:\ first, their optical characteristics can be designed to match the particular application, including operation at low-loss telecom wavelengths and matching resonances with other systems (e.g.\ atomic transitions). Second, the mechanical modes can be designed to coherently store quantum information for more than ten microseconds~\cite{Wallucks2020}, unparralelled for systems natively operating at telecom wavelength. Third, the mechanical mode offers a direct interface to other quantum systems operating in the gigahertz frequency regime~\cite{Forsch2020,Jiang2020,Mirhosseini2020}, such as superconducting qubits, or spin quantum systems.
 
 Quantum teleportation~\cite{Bennett1993} of an unknown input state from an outside source onto a quantum node is considered one of the key components of long distance quantum communication protocols~\cite{Briegel1998,Sangouard2011}. It has been demonstrated with pure photonic quantum systems~\cite{Bouwmeester1997,Furusawa1998,Ma2012,Valivarthi2016} as well as atomic~\cite{Olmschenk2009} and solid-state spin systems~\cite{Pfaff2014} linked by photonic channels. While quantum teleportation involving the vibrational modes of a diamond has previously been demonstrated~\cite{Hou2016}, the extremely short lifetimes of the system required the mechanical state to be measured before the teleportation protocol was completed. This reverse time-ordering, as well as the operation in the visible wavelength regime, makes the protocol unsuitable for long distance quantum communication.

Here we demonstrate for the first time quantum teleportation of an arbitrary input state onto a long-lived
optomechanical quantum memory. In particular, we teleport a polarization encoded photonic qubit at telecom wavelength onto a dual-rail encoded optomechanical quantum memory. The memory is composed of two mechanical resonators, where the quantum information is stored in the single-excitation subspace of the two resonators. The teleportation we perform implements all components of first-level entanglement swapping~\cite{Jiang2007,Sangouard2011}. Together with the remote generation of a single-excitation~\cite{Riedinger2018}, or DLCZ-type~\cite{Duan2001} entanglement, which has been shown individually before, this current experiment demonstrates the combined requirements for a fully functional quantum repeater node~\cite{Jiang2007}. Besides the impact this has on quantum technologies, it also opens the way to create single-phonon arbitrary qubit states of massive, mechanical oscillators, which can be used for testing quantum physics itself and potential decoherence mechanisms leading to quantum-to-classical transition~\cite{Bassi2013,Frowis2018}.

\begin{figure*}
\centering
\includegraphics[width=183mm]{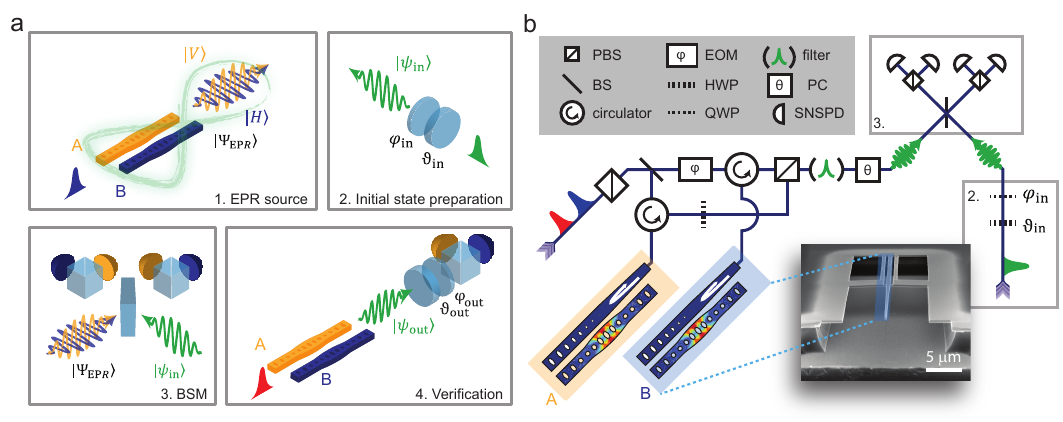}
    \caption{\textbf{Teleportation protocol and experimental setup.} a) Schematic representation of the key steps of the teleportation protocol and its verification. 1.\ Realization of an Einstein-Podolsky-Rosen (EPR) source:\ the Stokes scattering of a pair of nanobeams results in an entangled state between the photon polarization state and the phonon population state in the nanobeams. 2.\ An arbitrary input state is encoded in the polarization basis of a weak coherent state. 3.\ A Bell-state measurement (BSM) of the polarization teleports the input state onto the joint mechanical memory state. 4.\ A short anti-Stokes pulse maps the teleported state $\ket{\psi_\outp}$ back onto the photon polarization for verification. b) Schematic diagram of the experimental setup. Each nanobeam is placed in one of the arms of a phase-stabilized interferometer, where the polarization in one arm is rotated from horizontal to vertical using a half-wave plate (HWP), before recombining the paths on a polarizing beamsplitter (PBS). The control pulses are filtered out with a narrow linewidth Fabry-P\'{e}rot filter. The electro-optic-modulator (EOM) and Pockels cell (PC) allow fast selection of the readout basis in the verification step. Also shown are the half- and quarter-wave plate (QWP) used for the input state preparation, and BSM beamsplitter (BS), PBSs and superconducting nanowire single-photon detectors (SNSPD). (inset) Electron micrograph of one of the optomechanical devices used.}
    \label{Fig:1}
\end{figure*}

Our optomechanical register consists of two silicon photonic crystal nanobeams, A and B, on two separate chips~\cite{Riedinger2018}. Both the nanobeams support a co-localized optical and mechanical mode with resonance frequencies in the optical telecom C-band around \SI{1550}{nm} and the microwave C-band around \SI{5}{GHz}, respectively. The optical and mechanical modes are coupled through the radiation pressure force and photoelastic effect with a  single photon coupling rate $g_0/2\pi \approx \SI{900}{kHz}$. The chips are placed $\SI{20}{cm}$ apart from each other inside a dilution refrigerator, and the nanobeam resonators are cryogenically cooled close to their quantum ground state of motion. Optical control pulses of \SI{40}{ns} length that are either blue or red detuned by one mechanical frequency $\Omega\und{m}$ from the optical resonance give rise to linearized optomechanical interactions, addressing the Stokes and anti-Stokes transitions of the system, respectively~\cite{Aspelmeyer2014,Riedinger2016}.

Our teleportation protocol is based on the proposal described in Ref.~\cite{Li2020,Pautrel2020}, which is schematically shown in Fig.~\ref{Fig:1}a, while a sketch of the experimental setup can be seen in Fig.~\ref{Fig:1}b. Each optomechanical device is placed in one of the arms of an actively phase-stabilized fiber interferometer, see Supplementary Information (SI) section~\ref{phase_lock} for more details. The paths are recombined on a fiber-polarizing beamsplitter such that the photons from the devices are cross polarized. A single blue detuned pulse is injected into the interferometer exciting each nanobeam with the same probability $\pb$, and the which-path information of the Stokes scattered light is encoded in the polarization state of the optical mode. Light from device A is vertically, while light from device B is horizontally polarized. The joint state of the two mechanical resonators $\AB$ and the optical field "$\opt$" after recombining, can be described as

\begin{align}
    \nonumber\ket{\Psi_\mathrm{EPR}} &\propto \ket{0}_\opt \ket{00}_\AB \\\nonumber &\hspace{15pt}+ \sqrt{\pb} \left(\ket{H}_\opt \ket{01}_\AB + e^{i\phi}\ket{V}_\opt \ket{10}_\AB\right) \\ &\hspace{30pt} + \order{\pb}
    \label{Eq:1}
\end{align}
where $\ket{0},\ket{1}$ denote the number states containing 0 and 1 excitation, respectively, $\phi$ can be set by controlling the relative phase of the light coming from each nanobeam using an EOM in arm A of the interferometer, and $\pb \ll 1$ is the Stokes scattering probability set by the blue detuned control pulse energy. Conditioned on the presence of a Stokes-scattered photon, Eq.~\eqref{Eq:1} is the Einstein-Podolsky-Rosen (EPR)-state that forms our basic resource for teleportation (as shown in Fig.~\ref{Fig:1}a, top left). 

After passing a narrow-band ($\sim$$\SI{40}{MHz}$) Fabry-P\'{e}rot optical cavity filter to reject the excitation pulse light, the optical part of our EPR-state wave-packet is sent to a (potentially remote) Bell-state measurement apparatus. The arbitrary input qubit state, $\ket{\psi_\inp}$, to be teleported is encoded into the polarization of a weak coherent state obtained from a heavily attenuated independent laser (Fig.~\ref{Fig:1}a, top right)

\begin{equation}
    \ket{\psi_\inp} \propto \ket{0} + \alpha \left(\cos{\frac{\theta_\inp}{2}}\ket{H} + e^{i \phi_\inp}\sin{\frac{\theta_\inp}{2}} \ket{V}\right) + \order{|\alpha|^2}
    \label{Eq:2}
\end{equation}
where $|\alpha|$ is the coherent state amplitude, and input angles $\theta_\inp$, $\phi_\inp$ can be chosen by setting the appropriate angles $\vartheta_\inp$, $\varphi_\inp$ on the waveplates shown in Fig.~\ref{Fig:1}b. 

We then implement a polarization based Bell-state measurement (BSM) by combining $\ket{\psi_\inp}$ with the optical part of $\ket{\Psi_\mathrm{EPR}}$ on a 50/50 beamsplitter (Fig.~\ref{Fig:1}a, bottom left) and further analyzing the output polarization using polarizing beamsplitters and single-photon detectors. From Eq.~\eqref{Eq:1} and~\eqref{Eq:2} we can see that both the conditional EPR state and the photonic state to be teleported are close to the single excitation ideal case, and, by operating in the suitable regime $\sqrt{\pb} \ll |\alpha| \ll 1$, we can beat the classical threshold despite the higher order terms that reduce the teleportation fidelity (see Ref.~\cite{Li2020} and SI section~\ref{Simulations}). In this limit a coincidence between polarizations $H$ and $V$ in the BSM projects the state of the mechanical resonators onto

\begin{equation}
    \ket{\psi_\outp}_\AB = \cos{\frac{\theta_\inp}{2}}\ket{10}_\AB \pm e^{i \phi_\inp}\sin{\frac{\theta_\inp}{2}} \ket{01}_\AB
    \label{Eq:3}
\end{equation} 
where the $+$ ($-$) corresponds to cases where the coincidence occurred on the same (different) output port of the BSM beamsplitter. This event corresponds to the input state $\ket{\psi_\inp}$ being teleported onto the single-excitation subspace of the two mechanical resonators. We note that, as also stated in~\cite{Li2020}, the teleported mechanical state has the probability amplitudes of the two eigenstates exchanged (bit flip) and has a possible $\pi$-phase difference (phase flip) compared to the conditional EPR state of Eq~\eqref{Eq:1}. We take this into account in post-processing.

Finally, we can verify that the teleportation was successful by mapping the joint state of the mechanical resonators back onto an optical polarization state using a red detuned pulse (Fig.~\ref{Fig:1}a, bottom right). We can choose an arbitrary measurement basis for the polarization analysis setup by pulsing the EOM in arm A to adjust the relative phase $\phi_\ro$ between the $H$ and $V$ components, and setting the rotation of a Pockels cell (PC) pulsed at its half-wave-voltage to adjust $\theta_\ro$, the relative amplitude between the $H$ and $V$ components. Conditioned on a successful teleportation event, the fidelity of the teleported state can be measured from the number of readout events with the polarization equal to the one in the input, divided by the total number of readout events (see SI section~\ref{Table} for more).

\begin{figure}[h!]
\centering
\includegraphics[width=\linewidth]{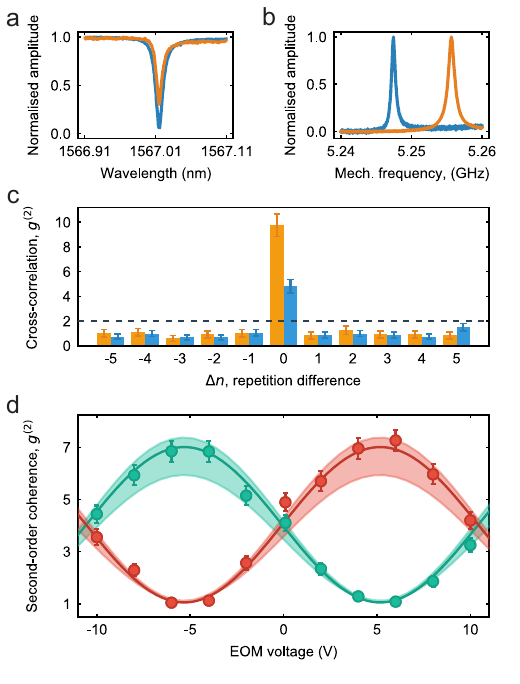}
	\caption{\textbf{EPR source characterization.} a) Characterization of the optical resonances of device A (blue) and device B (orange) measured in reflection. The devices have a small mismatch of about \SI{0.01}{GHz}. b) Mechanical spectra of the devices using optomechanical induced transparency (OMIT, see SI section~\ref{dev_fab} for more), device A in blue and device B in orange. The difference in the mechanical frequencies shown here, equal to \SI{8}{MHz}, is compensated by using a serrodyne-technique to frequency shift the light going to device A to ensure the emitted photons are fully indistinguishable. c) For the same repetition $\Delta n=0$, the Stokes and anti-Stokes fields of each devices show strong cross-correlations $g^{(2)}_\mathrm{cc}$, violating the bound for classical emitters (dashed line), while for different repetitions the detections are fully uncorrelated. Data in blue for device A and in orange for device B. d) Characterization of the entangled states produced by the EPR source:\ Second-order coherence $g^{(2)}_{i,j}$, in green for $i\neq j$ and in red for $i=j$, with $i,j \in \{D,A\}$ the detected polarization state of the Stokes and anti-Stokes photons respectively ($D$ for diagonal, $A$ for anti-diagonal, see main text), as a function of the phase shift induced in one arm of the interferometer by applying a pulsed voltage to the EOM. The solid line is a sinusoidal fit to guide the eyes. Shaded regions are expected values from simulation, see main text for details. All error bars are one standard deviation. } 
	\label{Fig:2}
\end{figure}

For the above protocol to work, the nanobeams must meet very stringent criteria. In particular, the emitted photons from each device must be completely indistinguishable in all degrees of freedom but their polarization, which in principle requires the nanobeams to have identical mechanical and optical resonance frequencies. In order to realize these requirements, we fabricate two chips with around 160 devices each and map their optical resonances, which results in about 40 matching pairs between the two chips in a single fabrication run (see SI section~\ref{dev_fab}).

A key challenge in the demonstration of quantum teleportation in our system is the rates at which we can detect threefold coincidences:\ the first two-photon coincidence to herald a teleportation event in the BSM and a subsequent readout photon to verify its success. While higher optical pulse energies result in higher scattering rates, and therefore higher coincidence rates for the teleportation protocol, they also increase the thermal population of the mechanical systems due to optical absorption heating~\cite{Meenehan2014} and lead to a reduced fidelity of the teleported state and its readout~\cite{Li2020}. We characterize the thermal population of the mechanical modes as a function of the optical pulse energy using sideband asymmetry~\cite{Riedinger2016}, and select a few pairs of devices that exhibit the highest scattering rates at a fixed total thermal population. A final selection criteria for the pair is their mechanical lifetime $T_1$, which should be long enough to store the teleported state until it is retrieved, however not too long, as the repetition rate of the experiment is set by the time required for the mechanical modes to re-thermalize into the ground state (see SI section~\ref{Simulations} for more details).

The characterization of the pair of devices we chose for this work is shown in Fig~\ref{Fig:2}. Their resonance wavelengths are $\lambda\und{c,A}= \SI{1567.0154}{nm}$ and $\lambda_\textrm{c,B}=\SI{1567.0153}{nm}$ with a difference in frequency of \SI{0.01}{GHz}, and with total optical linewidths $\kappa\und{A}/2\pi=\SI{1.47}{GHz}$ and $\kappa\und{B}/2\pi=\SI{1.18}{GHz}$ (intrinsic linewidths $\kappa\und{i,A}/2\pi=\SI{557}{MHz}$ and $\kappa\und{i,B}/2\pi=\SI{264}{MHz}$), see Fig~\ref{Fig:2}a. The small residual mismatch in mechanical frequencies $\Omega\und{m,A}/2\pi=\SI{5.2474}{GHz}$ and $\Omega\und{m,B}/2\pi=\SI{5.2555}{GHz}$ of about \SI{8}{MHz} (Fig~\ref{Fig:2}b) can be compensated by frequency shifting the optical excitation pulse in one of the arms, which we realize through serrodyning by driving an EOM with a linear ramp~\cite{Riedinger2018}. We measure $T_1^{(A)}=\SI{1.3}{\mu s}$, $T_1^{(B)}=\SI{1.9}{\mu s}$ such that we can operate our experiments with a $\SI{20}{\mu s}$ repetition time (see SI section~\ref{mech_and_heating} for details). We choose to work with scattering probabilities for the Stokes process of \SI{1.2}{\%} (\SI{1.3}{\%}) and for the anti-Stokes process of \SI{2.6}{\%} (\SI{2.9}{\%}) for device A (B), which correspond to a pulse energy of \SI{22}{fJ} (\SI{18}{fJ}) and \SI{50}{fJ} (\SI{40}{fJ}) for device A (B) (see SI sections~\ref{mech_and_heating} and~\ref{Simulations} for more details). 

We then confirm that the mechanical modes of the nanobeams can individually be prepared close to a single phonon Fock state. In order to accurately predict the $g^{(2)}$ of the mechanical state~\cite{Riedinger2016}, we measure the cross-correlation between Stokes and anti-Stokes scattered photons:\ $g^{(2)}_\mathrm{cc} = \frac{p_{\mathrm{S} \land \mathrm{aS}}(\Delta n)}{p_\mathrm{S} p_\mathrm{aS}}$, with $p_{\mathrm{S} \land \mathrm{aS}}(\Delta n)$ the probability to detect both a Stokes ($\mathrm{S}$) and anti-Stokes ($\mathrm{aS}$) scattered photon $\Delta n$ experimental repetitions apart and $p_\mathrm{S}$, $p_\mathrm{aS}$ the probability to detect individual Stokes and anti-Stokes photons, respectively. The measurement results shown in Fig.~\ref{Fig:2}c demonstrate cross-correlation values far above the classical bound of 2~\cite{Riedinger2016} by more than 9 (5) standard deviations for device A (B), proving that we can store our teleported state with little added noise. We also use this measurement to estimate the total thermal population added by the optical pulses. We infer a total thermal excitation of $\SI{0.24\pm0.04}{}$ ($\SI{0.10\pm0.01}{}$) for devices A (B) using a fixed delay between the pulses of \SI{100}{ns}~\cite{Hong2017}. Note that the average thermal excitation of the two devices is below $\lesssim0.24$, the threshold to demonstrate quantum teleportation~\cite{Li2020}.  

Having chosen the most suitable pair of optomechanical devices using the above criteria, we then proceed to characterize their suitability as a conditional EPR source to produce the state $\ket{\Psi_\mathrm{EPR}}$ in Eq.~\eqref{Eq:1}. We test the full conditional EPR source by proving the non separability of the Stokes field with the phononic state in the nanobeams, verifying it with the visibility after readout, similarly to~\cite{Riedinger2018}. We first measure the polarization of the optical output state of $\ket{\Psi_\mathrm{EPR}}$ in a rotated, diagonal basis. This projects the state of the nanobeams onto

\begin{equation}
   \ket{\psi}_\AB = \frac{1}{\sqrt{2}} \left(\ket{01}_\AB \pm e^{i \phi} \ket{10}_\AB\right)
   \label{Eq:4}
\end{equation}
where the sign $\pm$ depends on whether a diagonal $\ket{D}  \propto  \ket{H} + \ket{V}$ or anti-diagonal $\ket{A}  \propto  \ket{H} - \ket{V}$ photon was detected. We then map the joint state of the mechanical resonators back onto an optical polarization state, and measure it in the same diagonal basis as a function of $\phi$ set by the EOM voltage in arm A. This yields the second-order coherence $g^{(2)}_{i,j}(\phi) =  \frac{p_{\mathrm{S},i \land \mathrm{aS},j}}{p_{\mathrm{S},i} p_{\mathrm{aS},j}}$, with $i,j \in \{D,A\}$ the detected polarization state of the Stokes and anti-Stokes photons, respectively. As shown in Fig.~\ref{Fig:2}d, the cases where $i=j$ (the same polarization is detected for Stokes and anti-Stokes photons) exhibit opposite correlation compared to the cases where $i\neq j$. We obtain a visibility between Stokes and anti-Stokes pulses of $\mathcal{V} = \frac{g^{(2)}_{i=j} - g^{(2)}_{i \neq j} }{g^{(2)}_{i=j} + g^{(2)}_{i \neq j} } = \SI{74\pm3}{\%}$ for the EOM voltage of \SI{6}{V}, which shows the suitability of the optomechanical system as the EPR source for the teleportation. From this measurement we also calibrate our readout angle $\phi_\outp$ as a function of the applied EOM voltage. Finally, we compare the measured visibility to a numerical simulation of the entanglement experiment, taking as input only the independently measured thermal occupations and lifetimes of the two nanobeams (see SI section~\ref{mech_and_heating}), detector dark count probability, control pulse leakage and interferometer dephasing. The values expected from simulation are shown as the shaded region in Fig.~\ref{Fig:2}d, also taking into account the statistical  uncertainties in the measured simulation input parameters (see SI section~\ref{Simulations} for more details).

\begin{figure}
\centering
\includegraphics[width=\linewidth]{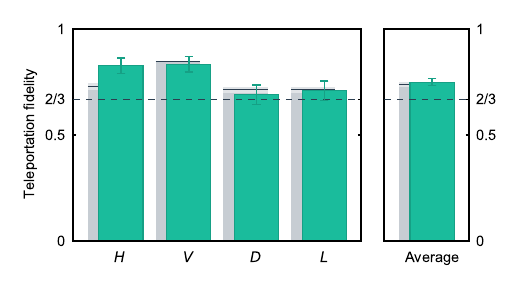}
    \caption{\textbf{Experimental quantum teleportation.} We measure the fidelity of the teleported state (green bars) for 4 different input basis states:\ Horizontal ($H$), Vertical ($V$), Diagonal ($D$) and circular left ($L$). The number of events for each basis, using a \SI{80}{ns} detection window, are $N_\mathrm{total} = 110,102,100, 99$ for $H$, $V$, $D$, $L$, respectively. All error bars are one standard deviation. The dashed line shows the classical limit of 2/3 and the rightmost bar the average $\expec{F} = 1/6( F_{H}+F_{V}+2 F_{D} + 2 F_{L}) = \SI{75.0\pm1.7}{\%}$. This demonstrates quantum teleportation beating the classical threshold by 4.8 standard deviations. The gray bars are the expected fidelities for each basis from our simulation of the full teleportation protocol, taking only independently measured parameters as input, with the shaded area the statistical confidence interval based on the uncertainty in those parameters and the black line the expected value.}
    \label{Fig:4}
\end{figure}

Since our scheme is symmetric around a change of $\pi$ in the phase of the teleported state, we verify our teleportation for input states $H$, $V$, $D=(H+V)$ and $L=(H+iV)$, only. To estimate our teleportation fidelity we set our EOM readout phase $\phi_\ro$ and Pockels cell angle such that the ideal teleported state is mapped to $H$ ($V$) for cases where the coincidence occurred on the same (different) output port of the BSM beamsplitter. We can then estimate the fidelity of the teleported state as the fraction of correct readout results $F_{i} = N_\mathrm{correct,i}/N_\mathrm{total,i}$ with $i\in \{H,V,D,L\}$ (see SI section~\ref{Table}). Note that this fidelity estimation includes the fidelity of our readout process, and therefore provides a lower bound to the true teleportation fidelity. After around 100 successful measurement runs for each basis we obtain an average fidelity of $\expec{F} = (F_{H} + F_{V} + 2(F_{D}+F_{L}))/6 = \SI{75.0\pm1.7}{\%}$ (see Fig.~\ref{Fig:4}), which is significantly above the classical threshold of 2/3 by 4.8 standard deviations. We obtain very similar results even when using the more robust Agresti-Coull interval~\cite{Brown2001}, which is above the threshold by more than four quantiles of a standard normal distribution. The average probability of the measured teleportation event is $\SI{4.3e-9}{}$ (one event every 1.3 hours).

Using our simulated version of the protocol, which agrees well with the measured fidelities, we can estimate the errors induced by various parts of the protocol. Excluding the readout error from the thermal population added by the red detuned pulse we estimate the teleportation fidelity to be $\expec{F}_\mathrm{sim} = \SI{77\pm1}{\%}$. By replacing the input weak coherent state with a true single photon state, our expected fidelity increases to $\SI{86\pm1}{\%}$, with all the other parameters unchanged.\\

Our experiment marks the first realization of quantum teleportation of an arbitrary qubit input state onto a dual-rail-encoded long-lived optomechanical quantum memory In contrast to previous experiments, our fully engineered system directly functions as a basic quantum repeater node for a future quantum network with an integrated memory component at telecom wavelengths. Even though the experiment was performed in the lab scale, in our current implementation we already separate the weak coherent state (and the BSM PBS) from the mechanical oscillators by several tens of meters of fiber. A logical further extension to the teleportation protocol will be to demonstrate a full first-level entanglement swapping operation~\cite{Jiang2007,Sangouard2011}, where previously distributed entangled states between distant nanobeams are combined to produce robust dual rail encoded entanglement. In the present form this would require four compatible nanobeam resonators. An attractive alternative would be to use nanobeams with two distinct frequency mechanical modes coupled to a single optical mode, combined with photonic dual rail frequency encoding. 

Our experiments also paves the way for transferring arbitrary qubit quantum states onto a mechanical system, which could lead to new tests of fundamental quantum physics~\cite{Aspelmeyer2014}. Another exciting possibility for our system is the potential to directly interface with various different quantum systems, such as superconducting microwave circuits, for example~\cite{Chu2020}. In this context the demonstrated teleportation could function as a quantum state transfer between photonic and microwave qubits. Moreover we show that the system can be easily mode matched to an outside source making it suitable to connect a large variety of optical systems that emit in the near-infrared.

\begin{acknowledgments}
We would like to thank K. Hammerer and R. Stockill for valuable discussions. This work is supported by the Foundation for Fundamental Research on Matter (FOM) Projectruimte grant (16PR1054), the European Research Council (ERC StG Strong-Q, 676842 and ERC CoG Q-ECHOS, 101001005), and by the Netherlands Organization for Scientific Research (NWO/OCW), as part of the Frontiers of Nanoscience program, as well as through Vidi (680-47-541/994) and Vrij Programma (680-92-18-04) grants. R.B.\ and T.A.\ acknowledge funding from the Funda\c{c}\~{a}o de Amparo \`{a} Pesquisa do Estado de S\~{a}o Paulo (2019/01402-1, 2016/18308-0, 2018/15580-6 and 2018/25339-4) and from the Coordena\c{c}\~{a}o de Aperfei\c{c}oamento de Pessoal de N\'{\i}vel Superior (Finance Code 001). B.H.\ also acknowledges funding from the European Union under a Marie Sk\l{}odowska-Curie COFUND fellowship.
\end{acknowledgments}

\textbf{Author Contributions:}\ N.F., B.H., A.W., J.L., and S.G. devised and planned the experiment. R.B. and B.H. fabricated the sample and N.F., B.H., R.B. and A.W. built the setup and performed the measurements. B.H. developed the code for the simulations. N.F., B.H., and S.G. analyzed the data and wrote the manuscript with input from all authors. T.P.M.A. and S.G. supervised the project.

\textbf{Competing Interests:}\ The authors declare no competing interests.

\textbf{Data Availability:}\ Source data for the plots are available on \href{https://doi.org/10.5281/zenodo.5079912}{Zenodo}.

\textbf{Code Availability:}\ The QuTiP code used for the simulations in the Supplementary Information is available on \href{https://github.com/GroeblacherLab/Optomechanical_Quantum_Teleportation}{GitHub}.

\setcounter{figure}{0}
\renewcommand{\thefigure}{S\arabic{figure}}
\setcounter{equation}{0}
\renewcommand{\theequation}{S\arabic{equation}}

\clearpage

\section*{Supplementary Information}

\subsubsection{Device fabrication and characterisation}
\label{dev_fab}
The nanobeams are fabricated from a silicon-on-onsulator (SOI) wafer with a \SI{250}{nm} device layer. We pattern our device structure using electron beam lithography, combined with HBr/Ar reactive ion etching. We perform a piranha cleaning step and finally release the device layer with a hydroflouric acid (HF) wet etch. To ensure that the two sets of devices are as identical as possible, we fabricate two lines of devices on the same chip which is then diced into half. The dicing also allows to couple to the devices waveguides with a lensed fiber (as done in~\cite{Hong2017}).

\begin{figure*}
	\includegraphics[width=0.5\linewidth]{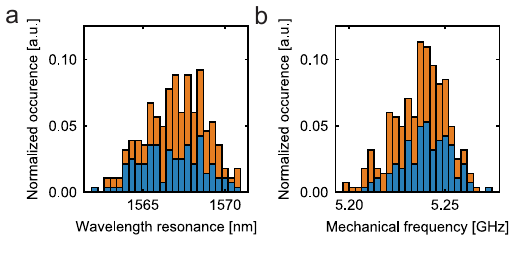}
	\caption{a) Histogram of the wavelength of the optical resonances from the two chips used in this experiment. The standard deviation of the distributions are \SI{1.9}{nm} and \SI{1.7}{nm} and the distance between the centers of the distributions is \SI{0.5}{nm}. b) Histogram of the mechanical resonance frequency. In this case, the standard deviation of the distributions are \SI{12}{MHz} and \SI{14}{MHz}, respectively, with a distance between the centers of the distributions of \SI{5}{MHz}. In both figures the chip with device A is in blue, the chip with device B is in orange and they have a total of 121 and 162 working devices respectively.  }
	\label{Fig:1_SI}
\end{figure*}

We proceed to characterize the optical and mechanical resonances of all the devices on the two chips. In order to measure the mechanical resonances, we detect the thermal motion of the nanobeam encoded in the amplitude noise of continuous wave (CW) light reflected from the device, using a fast detector connected to a spectrum analyzer. In Fig.~\ref{Fig:1_SI} it's shown the histogram of the optical (a) and mechanical (b) resonances of the two chips. The total spread of the optical cavity wavelengths is $\sim$\SI{1.8}{nm}, and $\sim$\SI{13}{MHz} for the mechanical center frequencies. The precision of the nanofabrication allows us to match pairs of nanobeams, that have small separation between the optical and mechanical resonances. Out of the approximately 160 devices on each chip we find 40 pairs that have an optical frequency difference of less than the typical linewidth of $\sim$\SI{1}{GHz}. The maximum frequency difference between the mechanical resonances is set by the maximum frequency that we can compensate by using a serrodyne technique (see section~\ref{setup} and main text for more). For the final pair chosen, we also perform a series of OMIT measurements varying the intracavity photon number, as described in~\cite{Safavi-Naeini2011}, from which we then calculate the $g_0$ of each devices.

\subsubsection{Mechanical lifetime and optical heating}
\label{mech_and_heating}

\begin{figure*}
	\includegraphics[width=0.5\linewidth]{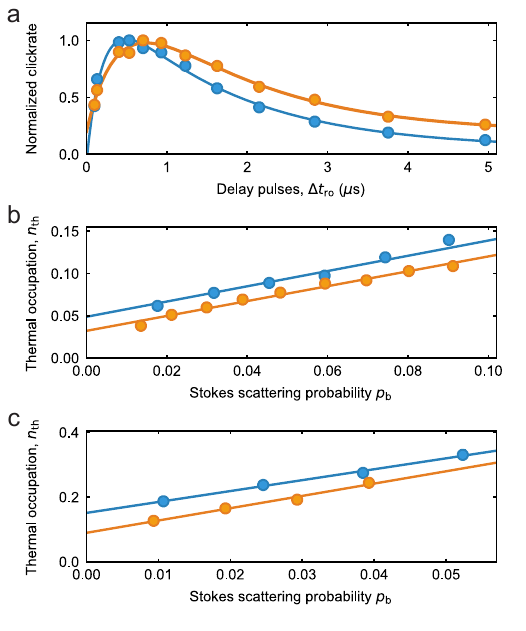}
	\caption{a) Normalized clickrates of a red detuned probe pulse delayed by $\Delta t_{\mathrm{ro}}$ from a heating pulse. The pump-probe scheme allows to track the changes in the thermal population (proportional to the clickrates) while sweeping the delay between the pulses~\cite{Hong2017}. The rise and decay of the thermal population fits well to a double-exponential phenomenological model, from which we obtain the $T_1$ of the devices ($\SI{1.3}{\mu s}$ for device A and $\SI{1.9}{\mu s}$ for device B). The raise times are $\SI{0.2}{\mu s}$ for device A and $\SI{0.4}{\mu s}$ for device B. Note that in the teleportation experiment we use a delay between the pulses of \SI{100}{ns} to limit the thermal heating. b) Thermal occupation $n_\mathrm{th}$, estimated from Stokes / anti-Stokes scattering rate asymmetry, as a function of the Stokes scattering rate $\pb$. The thermal occupation is increased due to the instantaneous optical absorption heating that occurs during the optical pulse. Solid line is a linear fit from which we find $\nth^{(\mathrm{A})} = n_0^{(\mathrm{A})} + \zeta_\mathrm{inst.}^{(\mathrm{A})} \times \pb^{(\mathrm{A})} = (\SI{0.049\pm0.004}{})+(\SI{0.91\pm0.05}{})\times \pb^{(\mathrm{A})}$, $\nth^{(\mathrm{B})} = n_0^{(\mathrm{B})} + \zeta_\mathrm{inst.}^{(\mathrm{B})} \times \pb^{(\mathrm{B})} = (\SI{0.032\pm0.002}{})+(\SI{0.88\pm0.05}{})\times \pb^{(\mathrm{B})}$, the uncertainties are from the fit. c) Thermal population due to the combined effect of a heating pulse and the probe pulses. The heating pulse occurs \SI{100}{ns} prior to the probe pulses. We vary the pulse energy of the heating pulse to assess the contribution of the delayed heating to the total thermal population. Solid line is a linear fit from which we find the slope $\zeta_\mathrm{delayed}^{(\mathrm{A})} = \SI{3.4\pm0.2}{} $ and $\zeta_\mathrm{delayed}^{(\mathrm{B})}= \SI{4.1\pm0.2}{}$, the uncertainties are from the fit. The fitted values are used in our simulation to assess optimal optical pulse powers, see SI section~\ref{Simulations} for details. In all figures the data from device A are in blue and from device B in orange.}
	\label{Fig:2_SI}
\end{figure*}

One of the main source of reduction of the teleportation fidelity is the thermal population of the mechanical mode caused by optical absorption heating (see e.g.~\cite{Meenehan2015}). Heating occurs directly, during the optical pulse, which we refer to as \emph{instantaneous heating}. Additionally, as visible from  Fig.~\ref{Fig:2_SI}a, the thermal population of the mode continues to increase beyond the duration of the optical pulse, which we refer to as \emph{delayed heating}. In our experiment the teleported state is retrieved after \SI{100}{ns} to minimize delayed heating as much as possible~\cite{Meenehan2015,Riedinger2016}. We intentionally choose a pair with shorter lifetime, such that our experiment allows for a high repetition rate. This is due to the fact that we have to wait for several mechanical lifetimes for the mechanical mode to thermalize back to its ground state. Fixing the delay between Stokes and anti-Stokes pules to \SI{100}{ns} we show in Fig.~\ref{Fig:3_SI}a that two devices with mechanical relaxation times of $\approx\SI{3}{\mu s}$ minimize the required measurement time to violate the non-classical threshold for teleportation.  

For our experiment we choose a pair of nanobeams that has the highest scattering probability while still resulting in a thermal occupancy small enough to perform quantum teleportation above the classical threshold. The thermal occupation versus scattering probability for the pair used is shown in Fig.~\ref{Fig:2_SI}b. This thermal occupancy reflects the mode heating that occurs during the Stokes optical pulse, the instantaneous heating. To take into account also the effect of the delayed heating, we mimic the real experimental conditions by sending to the devices a red detuned pulse \SI{100}{ns} before the pulses used to measure the thermal occupation. The first pulse is not for optomechanical interaction, but only for heating, while the second pulse is used to measure the thermal occupancy of the devices. The result can be seen in Fig.~\ref{Fig:2_SI}c, where we vary the energy of the pre-heating pulse, and show the corresponding Stokes scattering probability for that pulse energy. The thermal occupation here is given by the total effect of both pre-heating pulse and readout pulses, however the slope allows us to infer the thermal population added by the delayed heating only (see section~\ref{Simulations} for more details).

With the pulse energy used in this work (\SI{22}{fJ} (\SI{18}{fJ}) and \SI{50}{fJ} (\SI{40}{fJ}) for device A (B) and for the Stokes and anti-Stokes processes, respectively) we measure a thermal population given by the blue detuned Stokes pulse of $\SI{0.053\pm0.003}{}$ ($\SI{0.036\pm0.003}{}$) and for the red detuned anti-Stokes pulse (including delayed heating from the blue one) $\SI{0.135\pm0.005}{}$ ($\SI{0.066\pm0.005}{}$) for device A (B), respectively.

\subsubsection{Simulations}
\label{Simulations}
\begin{figure*}
	\includegraphics[width=1\linewidth]{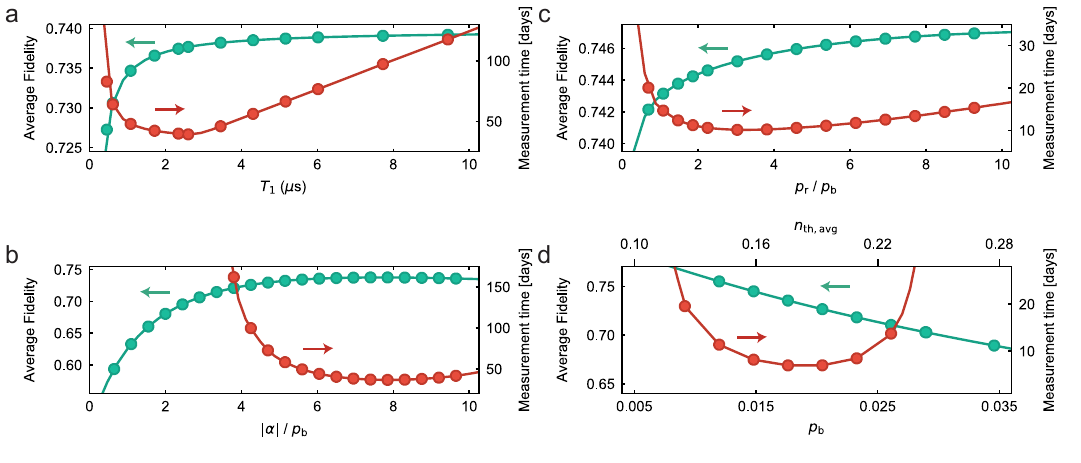}
	\caption{Average teleportation fidelity and total measurement time obtained from our simulations. The threshold for the measurement time is the required integration time for the fidelity to be above the threshold of 2/3 by at least 2 standard deviations for each basis. a) Varying the $T_1$ of the devices we see an optimal point at $T_1=\SI{3}{\mu s}$ (for simplicity both devices are assumed to have the same lifetime with a fixed readout delay of 100~ns). b) Varying the ratio between the coherent state amplitude $|\alpha|$ and the Stokes scattering probability $\pb$, we find an optimal value around 8. c) Varying the ratio between the Stokes and anti-Stokes processes, an optimal point is found for $\pr/\pb=3$. Here we use the parameters for instantaneous and delayed heating from the asymmetry measurements (Fig.~\ref{Fig:2_SI}b,c) and we choose the scattering rates to have a fixed total thermal population due to both pulses. d) Varying the scattering probability $\pb$, and so the average thermal occupation of the devices $n_{\mathrm{th,aveg}}$. Note a optimal point for $\pb = 0.02$. Also here the parameters for instantaneous and delayed heating from the asymmetry measurements are used. In all four figures the curves are guides to the eye.   }
	\label{Fig:3_SI}
\end{figure*}

We numerically simulate the cross-correlation, entanglement and teleportation experiments by keeping track of the total system density matrix throughout the respective protocols. We employ the QuTIP~\cite{Johansson2012,Johansson2013} simulation package to describe each subsystem in the Fock-basis up to $N=3$ excitation. We report that the calculated values differ less than \% from the results obtained with higher $N$ and have the advantage of a much shorter calculation time. For the teleportation experiment the total system is described by the state of the mechanical mode of each nanobeam, $\rho_{\mathrm{A}},\rho_{\mathrm{B}}$, the optical mode of the emitted light $\rho_H$, $\rho_V$ and the optical mode of the input state $\rho_H^\mathrm{in}$, $\rho_V^\mathrm{in}$. 

The teleportation model inputs include the input state angles $\vartheta_\inp$, $\varphi_\inp$, weak coherent state amplitude $|\alpha |$, Stokes scattering probability $\pb$, anti-Stokes scattering probability $\pr$ and readout angles $\vartheta_\ro$, $\phi_\ro$. The models for the cross-correlation and entanglement only take a subset of these.

We further include a number of imperfections as model inputs: 
\begin{itemize}
	
	\item The thermal occupation of the mechanical modes, consisting of three parts:\ First, a steady state thermal occupation of the mechanical mode $n_0^{(A,B)}$, likely caused by heating from the CW leakage from the pulse generation apparatus and heating from the average power sent to the nanobeam that is not efficiently dissipated. Second, the instantaneous heating due to the blue pump pulse. Third, the instantaneous heating due to the red pulse and the effect of the delayed heating from the blue pulse. We model the first two by taking the initial state of the mechanical modes of device A and B as $\rho_{\mathrm{A}}$, $\rho_{\mathrm{B}}$ at the beginning of the protocol to be in a thermal state with $\nth^{(\mathrm{A},\mathrm{B})} = n_0^{(\mathrm{A},\mathrm{B})} + \zeta_\mathrm{inst.}^{(\mathrm{A},\mathrm{B})} \times \pb^{(\mathrm{A},\mathrm{B})}$, where the $n_0^{(\mathrm{A},\mathrm{B})}$ and $\zeta_\mathrm{inst.}^{(\mathrm{A},\mathrm{B})}$ are estimated by the offset and slope of the fit in Fig.~\ref{Fig:2_SI}b, respectively. The third part is modeled by coupling the mechanical modes to a thermal bath with $\nth^{(\mathrm{A},\mathrm{B})} = \zeta_\mathrm{delayed}^{(\mathrm{A},\mathrm{B})} \times \pb^{(\mathrm{A},\mathrm{B})} + \zeta_\mathrm{inst.}^{(\mathrm{A},\mathrm{B})} \times (\pr^{(\mathrm{A},\mathrm{B})}-\pb^{(\mathrm{A},\mathrm{B})})$, where $\zeta_\mathrm{delayed}^{(\mathrm{A},\mathrm{B})}$ is estimated by the slope of the fit in Fig.~\ref{Fig:2_SI}c. The coupling to the thermal bath is modeled to occur just after the detection of the Stokes-scattered photon and the coupling is implemented by a two-mode-squeezing interaction with an angle $\sqrt{\nth}$ between each mechanical mode and an environmental vacuum mode, and subsequently tracing out the environmental mode. Using the fixed red and blue pulse energies used in the cross-correlation, entanglement and teleportation experiment, we extract the $\nth^{(\mathrm{A},\mathrm{B})}$, used in the simulation shown in Fig.~\ref{Fig:2}d and Fig.~\ref{Fig:4}, from the measured cross-correlation data (Fig.~\ref{Fig:2}c), as it is increased due to the additional average power sent to the device necessary for locking the devices interferometer. The statistical uncertainties of the estimated thermal occupations result in a simulated range of values, shown in the figures as shaded areas.
	
	\item The mechanical lifetime of each mechanical mode, modeled as an amplitude damping channel acting on each mechanical mode with loss probability $p_\mathrm{loss}^{(\mathrm{A},\mathrm{B})}=e^{-T_1^{(\mathrm{A},\mathrm{B})}/\Delta t_\mathrm{ro}}$, where $\Delta t_{\mathrm{ro}}= \SI{100}{ns}$ is the readout delay and $T_1^{(\mathrm{A},\mathrm{B})}$ as found in Fig.~\ref{Fig:2_SI}a.
	
	\item Detector dark-counts and excitation pulse leakage, resulting in a detector click not originating from either nanobeam or input photon. We take the weighted sum of the density matrix corresponding to a genuine coincidence click from a Stokes scattered photon and input photon, and the density matrix where one of the clicks was produced by a detector dark count or leakage photon. The weights take into account the darkcount and leakage click probability $p_\mathrm{dc}$, and the total Stokes photon detection efficiency (device coupling efficiency, detection filter and setup efficiency and single photon detector efficiency), see section~\ref{setup}.
	
	\item Finite interferometer phase stability, modeled as a dephasing channel with dephasing probability $V_\mathrm{int}$ on the optical mode $H$, see Fig.~\ref{Fig:1_SI}.
	
\end{itemize}

By including optical losses we can also estimate the protocols success probability. We use this simulation to obtain ideal device parameters and power settings to demonstrate teleportation, see Fig.~\ref{Fig:3_SI}. Firstly, we determined the optimal $T_1$ of the devices, a tradeoff between efficiently storing the state and having a higher repetition rate of the experiment. Secondly, we determine the optimal ratio of the weak coherent state (WCS) amplitude $|\alpha |$ and the scattering probability of the Stokes process $\pb$, given by the requirement that $|\alpha|$ has to be much larger than $\pb$ for the protocol to work, as well as much smaller than $1$ to avoid excessive contribution of the higher order terms of the coherent state. Note that, with a single photon source input, the simulated average fidelity of the teleportation increases to \SI{86}{\%} with the same parameters used in this work. We then calculate the best ratio of scattering probability of the Stokes and anti-Stokes processes, $\pr / \pb$. Here the optimal point is given by a tradeoff between having a high excitation probability of the devices, given by $\pb$, and a efficient readout of the mechanical state, given by $\pr$, while having a constant total thermal occupation. %We find an optimal ratio $3$. Note that the Stokes probability changes also the WCS amplitude and since this simulation relies on the less precise thermal calibration shown in~\ref{Fig:2_SI}c, the absolute estimation of the measurement time is not fully consistent with the previous results, however, gives a similar trend for a large range of parameters hence validating the result.
Finally, we sweep the Stokes scattering probability $\pb$ using the optimal settings of the other parameters. This also change the total thermal population, and so the average thermal population of the devices ($n_{\mathrm{th,avg}}$). The optimal point is for $\pb$ high enough to have high rates and with a thermal population low enough to have a average fidelity of the teleported states above the threshold.

\subsubsection{Optical setup and experimental procedures}\label{setup}

\begin{figure*}
	\includegraphics[width=\linewidth]{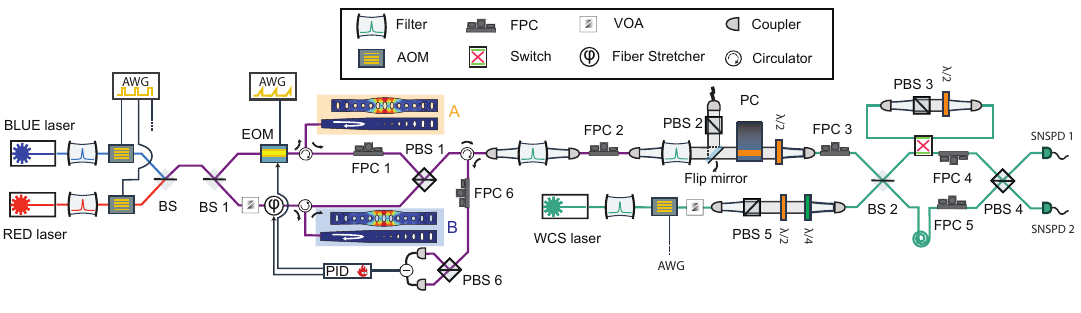}
	\caption{Detailed schematics of the setup (see the text for more details). AOM are the acousto-optic modulators, AWG the arbitrary waveform generators, EOM the electrooptic modulator, VOA the variable optical attenuator, BS the 50/50 beamsplitters, PBS the polarization beamsplitters, PC the Pockels cell, FPC the fiber polarization controller, $\lambda/2$ half-wave plate, $\lambda/4$ quarter-wave plate and SNSPD the superconducting nanowire single-photon detectors. WCS is the weak coherent state to be teleported. Note that PBS 2, 3 and 5 are free space PBS while the others are fiber PBS.}
	\label{Fig:4_SI}
\end{figure*}

A complete schematic of the setup used in the experiment can be found in Fig~\ref{Fig:4_SI}. The optical pulses are generated from three separate tunable CW diode lasers, two for the optical control pulses (RED and BLUE lasers), one for the state to be teleported (WCS laser). The lasers are stabilized using a wavelength meter and a frequency beat-lock system (not shown), which allows to have a laser frequency jitter of few MHz. All light sources are filtered with fiber filters of \SI{50}{MHz} linewidth to reduce the amount of classical laser noise at GHz frequencies. The pulses used in the experiment are generated using two \SI{110}{MHz} acousto-optic modulators (AOMs) gated via an arbitrary waveform generator (AWG, Agilent 81180A) and are fed into the Mach-Zehnder interferometer through a fiber beamsplitter (BS 1). A variable optical attenuator (VOA) is used to have different pulse powers in the two arms of the interferometer. An electro-optical modulator (EOM) is used for the serrodyne shift (signal from an AWG, Tektronic AFG3152C) and, together with a home-built fiber stretcher, for the phase stabilization of the interferometer (see section~\ref{phase_lock} for more). The pulses are sent to the devices inside the dilution refrigerator via optical circulators, using lensed fibers to couple to the devices' waveguides. The light coming from the devices is recombined on PBS 1 and filtered with two free-space Fabry-P\'{e}rot with \SI{40}{MHz} total linewidth and \SI{17}{GHz} and \SI{19}{GHz} free spectral range. The average suppression ratio is \SI{4.0e-11}{} (\SI{104}{dB}) for the Stokes and \SI{1.5e-11}{} (\SI{108}{dB}) for the anti-Stokes process, where the difference in suppression arises from higher order modes present in the transmission spectra of the cavities due to a small mode mismatch. Every $\sim \SI{6}{s}$ the measurements are paused and CW light is injected in the filter cavities to lock them using optical switches (not shown, see~\cite{Riedinger2016}). A flip mirror and a free space PBS (PBS 2) are used to align the polarization to the axis of the Pockels cell (PC) using CW light. The signal from the nanobeams, once passed through the PC, is combined in a fiber 50/50 BS with the WCS and then routed to a fiber PBS (PBS 4) with a relative delay in arrival time of \SI{300}{ns}. An optical switch is used to send optical pulses on a free space setup ($\lambda /2$ and PBS 3) to analyze the polarization in order to align the polarization of the devices and the coherent state. The $\lambda /2$ after the PC is used for the test of the EPR source (Fig.~\ref{Fig:2}c) to change the polarization of the photon from the devices from the alignment settings ($H$, $V$) to the measurement setting ($D$, $A$).

We use fiber polarization controllers (FPC) to:
\begin{itemize}
	\item Maximize the transmission through the various PBSs along the setup (FPC 1 and othera that are not shown).
	\item Align the polarization to the axis of the PC using a reference PBS (PBS 2 and FPC 2).
	\item Align the polarization of the light from the device A and from the WCS with respect to a reference free-space PBS (PBS 3 and FPC 3), in both the $H$ and $D$ polarization.
	\item Align the polarization of the light from the device A and from the WCS with respect to the BSM PBS (PBS 4 and FPC 4, 5).
	\item Ensure that the relative phase between the devices on which we lock the interferometer is constant during the whole experiment (PBS 6 and FPC 6, see section~\ref{phase_lock} for more).
\end{itemize} 
Note that the polarization of device B is always orthogonal since the paths are recombined on PBS 1. In particular the alignment on the free-space PBS 3 is fundamental to determine the relative phase between the state of the devices and the teleported one, in order to properly map the Poincar\'e spheres of both states. 

To guarantee that the EPR state is the desired maximally entangled state, we assure that the measured scattering rates from the Stokes process of the devices are equal. Since for both devices the collection efficiency of the scattered photons and the thermal populations are similar, matching the measured rates means matching $\pb$. Note that the readout probability will be, in theory, identical for both devices while the scattering rates from the anti-Stokes process are not matched since they are sensitive to the (different) thermal occupancy of the devices. From the results of Fig.~\ref{Fig:3_SI}, we also calibrate the clickrates from the WCS to be 6.8 times the clickrates from the Stokes process of the single device.

The total measurement time of the experiment in the main text, Fig.~\ref{Fig:4}, was 35 days. To ensure the stability of the setup and to exclude data recorded during a faulty state of the setup, we check the parameters of the setup (the laser locking, the filter detection locking among others) every five seconds and we measure the polarization state and the clickrates from the devices and the WCS every hour. We further align the polarization as described above every 2 days, to keep the deviation in polarization and clickrates to a few \% and the decrease in visibility of the interferometer in the fixed EOM voltage by a few \% (see section~\ref{phase_lock} for more). We also align the free space optics every two weeks when the efficiency of the setup (measured via the clickrates at a fixed pulse energy) reduces by a few \%. The polarization check, which is performed every hour, requires approximately 1/6 of the total measurement time. This time together with the time in which the detection filters are re-locked (1/6, approximately 1 second every 6 seconds), brings to the total effective integration time to 23 days.

The average collection efficiency of the scattering photon from the devices (including the coupling from the fiber to the optical waveguide, equal to \SI{46}{\%} (\SI{55}{\%})) is \SI{11}{\%} (\SI{13}{\%}) for line A (B). The detectors dark-counts and residual leakage of the pump pulses are \SI{5.9e-6}{} (\SI{4.9e-6}{}) for the Stokes process with pulse energy of \SI{22}{fJ} (\SI{18}{fJ}) and \SI{6.7e-6}{} (\SI{5.6e-6}{}) for the anti-Stokes process with pulse energy \SI{50}{fJ} (\SI{40}{fJ}) for device A (B). We use for the simulation the average value, $p_{\mathrm{dc}} = \SI{6e-6}{}$. Note that the bare dark-counts from the SNSPDs are sub-Hz (\SI{0.1}{Hz}).

\subsubsection{Phase locking}\label{phase_lock}

\begin{figure*}
	\includegraphics[width=0.5\linewidth]{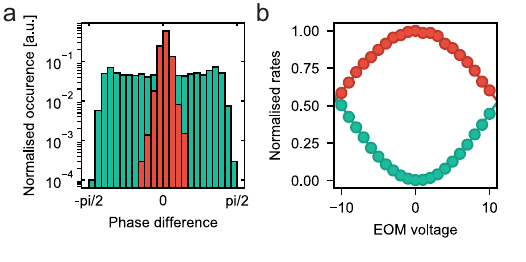}
	\caption{a) Occurrence histogram of the difference of the phase in the interferometer unlocked (green) and locked (red) configuration, for each dataset we sample the phase every second for approximately an hour. For the latter case we obtain a $\textrm{FWHM}\approx\pi/40$. b) Normalized count rates from the first order interference of a laser on resonance with the cavities and the detection filters for detector 1 (red) and detector 2 (green). Sweeping the EOM voltage in one of the interferometer arm shows the interference pattern. Due to the limited voltage range only half a period is visible. The interferometer visibility in this case is $V_\mathrm{int}\approx99.3\%$. Error bars are one standard deviation and are too small to be seen.}
	\label{Fig:5_SI}
\end{figure*}

In order to have a well-defined phase relation between the two nanobeams, so to have a stable $\phi$ in the EPR state (see Eq.~\ref{Eq:1}), we need a phase stabilized interferometer, which has a free-spectral range (FSR) of more than \SI{6}{GHz}. The red detuned locking pulse, delayed by $\SI{1}{\mu s}$ in time with respect to the measurement pulse, is reflected by the first cavity, split by a fiber PBS (PBS 6) and measured on a balanced photodiode. A PID program on a RedPitaya~\cite{Luda2019} is used to actively lock the interferometer during the experiment. The locking feedback has a bandwidth of $\sim\SI{3.8}{MHz}$ and two output signals, see Fig.~\ref{Fig:4_SI}:\ one is applied to the fiber stretcher that is low-passed by a pre-amplifier (\SI{30}{kHz}), while the other output is applied to the EOM in the arm of device A and its feedback is ultimately limited by the repetition rate of the experiment (\SI{100}{kHz}). In this way, we can compensate the high frequency components of the error signal (mainly given by the vibration picked up by the fiber in the dilution refrigerator induced by the pulsetube unit).

We directly estimate the phase stability from the normalized lock pulse amplitude, shown in Fig.~\ref{Fig:5_SI}a, obtaining a full-width at half maximum $\textrm{FWHM}\approx\pi/40$. We can also measure the long term phase stability using the average interferometer visibility during the whole measurement ($V_\mathrm{int}\sim99.3$\%, ultimately also limited by the polarization alignment). A typical measure of the first order interference from the laser light reflected from the devices is plotted in Fig.~\ref{Fig:5_SI}b, from which we can estimate a phase stability of $\pi/25$~\cite{Minar2008}. 

Note that the locking light and the signal light have two different paths, so we need to reference the relative phase between the devices in order to fix the EOM calibration done with the test of the EPR source (shown in Fig.~\ref{Fig:2}c). To do this we use a FPC (FPC 6) to maximize the interference visibility for a chosen EOM voltage, fixing the phase difference between the devices for every run of the experiment. Polarization drifts changes the point of maximum visibility over time. This because a change in polarisation after the PBS 1 changes the phase at which we lock the interferometer and so changes the relative phase between device A and device B. So, in order to have a small reduction in interferometer visibility for the chosen EOM voltage, the process of alignment is repeated every two days.

\subsubsection{Time evolution of the entangled state}\label{time_sweep}

\begin{figure*}
	\includegraphics[width=0.5\linewidth]{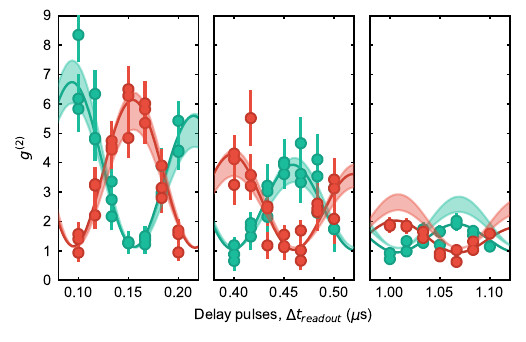}
	\caption{Second-order coherence $g^{(2)}_{i,j}$ for $i\neq j$ (green) and $i=j$ (red), with $i,j \in \{D,A\}$ the detected polarization state of the Stokes and anti-Stokes photons respectively (both occurring in the same repetition), as a function of the delay between the blue and red detuned pulses. The phase shift is induced by the difference in mechanical frequency of the two oscillators. We measure a decay time of $\SI{500}{ns}$. The shaded regions are the expected values from simulation, including the statistical uncertainties of the input parameters (see the text and section~\ref{Simulations} for more details). The solid lines are a fit to help guide the eye. All error bars are one standard deviation.}
	\label{Fig:6_SI}
\end{figure*}

To ensure that the quantum state teleported onto the joint mechanical state of the nanobeams maintains its coherence before the readout, we perform a characterization of the entangled state produced by the EPR source by sweeping the delay between the blue and red detuned pulses with a fixed offset in the EOM voltage, as previously done in~\cite{Riedinger2018}. The result is shown in Fig~\ref{Fig:6_SI}. We find a coherence time of the entangled state of $\SI{500}{ns}$, much longer than the readout delay $\Delta t_{\mathrm{ro}} = \SI{100}{ns}$. Note that this value is smaller than the measured $T_1$ of the single devices, which is a combined effect of additional delayed optical heating and dephasing. We compare the measured data to the simulation, as done in Fig~\ref{Fig:2}d, including the expected increase in thermal population due the the delayed absorption, using the fitted value of Fig~\ref{Fig:2_SI}b. We find that the reduction in coherence time can be explained by the delayed heating for short delays (up to $\approx\SI{500}{ns})$. For longer delay the small discrepancies can be caused by the imperfection in the simple model of the delayed heating.

\subsubsection{Hong-Ou-Mandel interference}\label{HOM}

\begin{figure*}[!]
	\centering
	\includegraphics[width=0.5\linewidth]{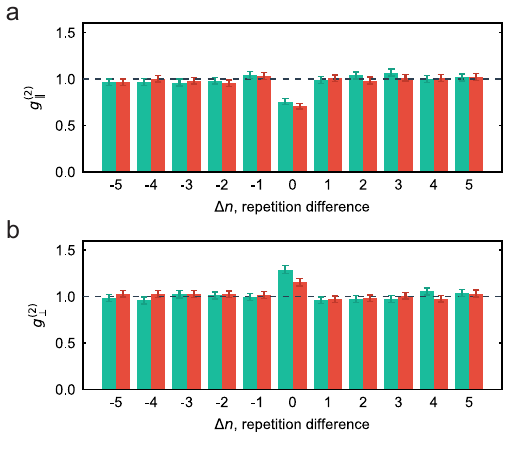}
	\caption{Two-photon quantum interference of nanobeam and input state photons. a) Second order correlation of the optical field from device A and the WCS (red) and device B and the WCS (green) for parallel polarization. A clear dip is visible for the same repetition. b) Same as a) for orthogonal polarisation. In this case a peak is visible for the same repetition. From this two dataset we can obtain a visibility of $V_{\mathrm{HOM}} = \SI{38\pm3}{\%}$ for device A and $V_{\mathrm{HOM}} =\SI{42\pm3}{\%}$ for device B. All error bars are one standard deviation.}
	\label{Fig:7_SI}
\end{figure*}

In order to further verify that the input photons from $\ket{\psi_\inp}$ and the photons scattered from the nanobeams in $\ket{\Psi_\mathrm{EPR}}$ are mode matched, as required for a successful BSM, we perform a Hong-Ou-Mandel (HOM) interference experiment~\cite{Hong1987}. We report in Fig~\ref{Fig:7_SI} the result for different repetition $\Delta n$ of the second order correlation of the optical field from the device A and the WCS (in red) and device B and the WCS (in green), for the case of interfering signals ($g^{(2)}_{\parallel}$) and not interfering signals ($g^{(2)}_{\perp}$). The coincidence rate from the not interfering signals (orthogonal polarization) is used for normalization of the HOM. The measured interference visibility of $V_{\mathrm{HOM}} =\frac{g^{(2)}_\parallel}{g^{(2)}_\parallel+g^{(2)}_\perp}$ is $\SI{38\pm3}{\%}$ for device A and $\SI{42\pm3}{\%}$ for device B. This results are consistent with HOM interference of ideal indistinguishable photons in a weak thermal and coherent state, for which a visibility of 2/5 = 40\% is expected. For different repetitions we find an average of $g^{(2)} = 0.986\pm0.01$ (device A) and $g^{(2)} = 1.007\pm0.017$ (device B).

\newpage

\subsubsection{Teleportation truth table}\label{Table}

We report the truth table used for the post-processing analysis, Tab.~\ref{Tab:1}. From Fig.~\ref{Fig:1}a (bottom left) we define the BSM detectors as $H_{\mathrm{early}}$, $V_{\mathrm{early}}$ and $V_{\mathrm{late}}$, $H_{\mathrm{late}}$, respectively from the leftmost to the rightmost. A successful teleportation event is a double click with opposite polarization. Note that, as visible in Fig.~\ref{Fig:4_SI}, this choice of name is given by the fact that in the experiment only two SNSPDs were used so that "early" or "late" refers to the output of the BSM BS (BS 2). For this reason the SNSPD 1 correspond to $H_{\mathrm{early}}$, $V_{\mathrm{late}}$ and SNSPD 2 to $V_{\mathrm{early}}$, $H_{\mathrm{late}}$. With the readout pulse, we verify the polarization of the teleported state choosing the appropriate measurement basis for the polarization analysis setup. We then compare the output of the measurement to the truth table to asses if the readout is correct.

\begin{table}[!h]
	\label{Tab:1}
	\begin{center}
		{\renewcommand{\arraystretch}{1.4}
			\begin{tabular}{| c | c | c | c | c |}
				\hline
				\textbf{Input polarization} & \multicolumn{4}{ c |}{\textbf{Teleportation event}}  \\ 
				\cline{2-5}
				& $H_{\mathrm{early}}$, $V_{\mathrm{early}}$ & $H_{\mathrm{early}}$, $V_{\mathrm{late}}$ & $H_{\mathrm{late}}$, $V_{\mathrm{early}}$  & $H_{\mathrm{late}}$, $V_{\mathrm{late}}$ \\
				\hline
				
				$H$ & $V$  & $V$  &$V$   & $V$  \\ \hline
				$V$ & $H$ & $H$  & $H$  &$H$ \\ \hline
				$D$  & $H$  & $V$ & $V$ &$H$ \\ \hline
				$L$  & $V$ & $H$ &$H$  &$V$ \\ \hline
				
			\end{tabular}
			\caption{Truth table used in the post-processing analysis of the data. }
		}
	\end{center}
\end{table}

\end{document}